\documentclass[
 reprint,
 amsmath,
 pre,
 amssymb,
 nofootinbib,
 aps,prl]{revtex4-1}

\usepackage[pdftex]{graphicx}
\usepackage{wasysym}
\usepackage{amsfonts}
\usepackage{ifsym}
\usepackage{hyperref} 
\usepackage{slashed}

\usepackage[bb=boondox]{mathalfa}

\def\l{\left(}
\def\r{\right)}

\def\sign{\mbox{sign\,}}

\newcommand{\be}{\begin{equation}}
\newcommand{\ee}{\end{equation}}
\newcommand{\bea}{\begin{eqnarray}}
\newcommand{\eea}{\end{eqnarray}}
\newcommand{\bg}{\begin{gather}}
\newcommand{\eg}{\end{gather}}
\newcommand{\bseq}{\begin{subequations}}
\newcommand{\eseq}{\end{subequations}}

\newcommand{\Tr}{{\rm Tr}}

\begin{document}
\title{
%From Confining Strings Towards Perturbative QCD 
 Confining Strings, Infinite Statistics and Integrability
 }
\author{John C. Donahue}
\author{ 
Sergei Dubovsky
}
\affiliation{
  Center for Cosmology and Particle Physics, Department of Physics,
      New York University,
      New York, NY, 10003, USA
      }

\begin{abstract}
We study confining strings in massive adjoint two-dimensional chromodynamics. Off-shell,  as a consequence of zigzag formation,
the resulting worldsheet theory provides a non-trivial dynamical realization of infinite quon statistics. Taking the high energy limit we identify a 
remarkably simple and novel integrable relativistic $N$-body system. Its symmetry algebra contains an additional ``shadow" Poincar\'e subalgebra.
This model describes the $N$-particle subsector of a $T\bar T$-deformed massless fermion.
\end{abstract}

\maketitle
%\section{Introduction}
{\it Introduction.}
$SU(N_c)$ quantum chromodynamics (QCD)  turns into a free string theory
in the planar ($N_c\to\infty$) limit \cite{'tHooft:1973jz}. 
For its maximally supersymmetric cousin the 
corresponding string theory has been identified as type IIB critical superstrings  on  $AdS_5\times S_5$  \cite{Maldacena:1997re}. 
Moreover, the corresponding worldsheet theory is integrable and has been solved, resulting in the exact spectrum of planar ${\cal N}=4$  Yang--Mills (YM) (see \cite{Beisert:2010jr} for an overview).

It has proven excruciatingly difficult to reproduce this success for planar non-supersymmetric gluodynamics.
%\footnote{Recall that in the planar limit quarks decouple from gluons, so that the full QCD reduces to
%the pure YM theory.}. 
The corresponding string theory has not even been identified yet, despite a  rich 45 year-long history of study.
One may suspect at first that the question is not sufficiently well-posed. However, a sharp version of the problem can be formulated as follows \cite{Dubovsky:2015zey}. Consider a background of YM theory with a single infinitely long confining string (flux tube). In the strict planar limit the string excitations decouple from bulk degrees of freedom and  give rise to a microscopic two-dimensional model. The challenge is to build this worldsheet theory.
 
 %For such a challenging problem, identifying the reason why it is so hard is already a step forward.
 With  this sharp formulation at hand, one  immediately understands the source of the difficulties. 
%Namely, lattice YM simulations  \cite{Athenodorou:2010cs,Athenodorou:2011rx,Athenodorou:2013ioa,Athenodorou:2016kpd,Athenodorou:2016ebg,Athenodorou:2017cmw} (see
%\cite{Teper:2009uf,Lucini:2012gg} for reviews)  exclude the presence of additional massless excitations on the worldsheet. 
%This implies that  the flux tube theory in $D=4$-dimensional space-time has  irreducible universal particle production \cite{Dubovsky:2012sh,Cooper:2014noa}, associated with the Polchinski--Strominger term \cite{Polchinski:1991ax}. 
%
Namely, in the absence of  additional massless degrees of freedom on the worldsheet, the flux tube theory in $D=4$-dimensional space-time has  irreducible particle production \cite{Dubovsky:2012sh,Cooper:2014noa}, associated with the Polchinski--Strominger term \cite{Polchinski:1991ax}. Lattice YM simulations  \cite{Athenodorou:2010cs,Athenodorou:2011rx,Athenodorou:2013ioa,Athenodorou:2016kpd,Athenodorou:2016ebg,Athenodorou:2017cmw} (see
\cite{Teper:2009uf,Lucini:2012gg} for reviews) rule out the presence of additional massless excitations on the worldsheet, thus excluding integrability. 
 
 Note that at $D=3$ integrability does not require additional degrees of freedom~\cite{Dubovsky:2015zey}. However, a closer look at the lattice data shows that the worldsheet theory is not integrable in $D=3$ YM either \cite{Dubovsky:2014fma,Dubovsky:2016cog,Chen:2018keo}.
 
 Hence, unlike for the ${\cal N}=4$ case, in the planar limit of non-supersymmetric YM
 we are left with an interacting non-integrable two-dimensional model.
 To make life harder (and more interesting),  at high energies this model exhibits  characteristically  gravitational behavior instead of that of a conventional quantum field theory \cite{Dubovsky:2012wk,Dubovsky:2018dlk}.
 
 This leaves two  directions for  further progress. First,  as gluodynamics does not have any relevant deformations, the worldsheet theory is likely isolated. This turns it into a natural target for the modern $S$-matrix bootstrap \cite{Paulos:2016fap,Paulos:2016but}, following the success of the conformal bootstrap in describing another isolated theory---the $3D$ Ising model \cite{Poland:2018epd}. A first promising step in this direction has been made very recently \cite{EliasMiro:2019kyf}.
 
 The other possibility is to identify a nearby integrable model, and to obtain a description of the worldsheet dynamics by performing a systematic perturbative expansion around this integrable theory.
 The quality of the latter expansion will then provide a natural measure of the proximity of the integrable mode to the full worldsheet theory.
 
  One may object that there is no {\it a priori} reason for a controlled integrable approximation to exist.
 However, the analysis of lattice data provides a number of tantalizing hints supporting this program. This motivated a proposal for an integrable 
 approximation---the Axionic String Ansatz (ASA)---both at $D=3$ and $D=4$ \cite{Dubovsky:2013gi,Dubovsky:2015zey,Dubovsky:2016cog,Donahue:2018bch}. 
   
   The successes of  the ASA have a clear physical origin \cite{Dubovsky:2018dlk,Dubovsky:2018vde}. At low energies the worldsheet degrees of freedom are translational Goldstone bosons of the non-linearly realized Poincar\'e symmetry \cite{Luscher:1980ac,Luscher:2004ib,Aharony:2010db,Aharony:2010cx,Dubovsky:2012sh,Aharony:2013ipa}. Their low energy dynamics is well-approximated by a classically integrable Nambu--Goto action.
   On the other hand, at high energies worldsheet degrees of freedom correspond to partons of perturbative QCD  (gluons in the YM case).
   Asymptotic freedom then implies that hard particle production is suppressed also at high energies. 
  
   This reasoning suggests that the violation of worldsheet integrability  may be a transient phenomenon dominated by intermediate  energies, $E\sim \Lambda_{QCD}$. 
   A combination of low  and high energy expansions may then allow one to describe the worldsheet dynamics at all scales. 
   
   As argued in \cite{Dubovsky:2018dlk}, a natural playground to test these ideas is provided by adjoint $D=2$ QCD. 
   As the  worldsheet theory lives in two dimensions for any $D$, general lessons from the study of the worldsheet should be universally applicable.   
   
%  Indeed, unlike for other questions (such as the glueball spectrum), it is natural to expect that general lessons learnt from the study of the worldsheet dynamics should be straightforward to transfer between theories in different numbers of space-time dimensions, given that the worldsheet always 
%  stays two-dimensional independently of $D$.
  
  The goal of the present paper is to report a solution to the very first step in this program---the identification of an integrable approximation at $D=2$.

%\section{Hamiltonian Formalism and Infinite Statistics}
%\label{sec:inf}
{\it Hamiltonian Formalism and Infinite Statistics.} 
Massive adjoint QCD$_2$ is defined by the following action
\be
\label{action}
S=\int d\tau d\sigma\Tr\l {-1\over 2 g^2} F_{\mu\nu}F^{\mu\nu}+\bar\psi(i\slashed\nabla^{(ad)}-m)\psi\r\;,
\ee
where $\psi$ is a Majorana fermion in the adjoint representation of the $SU(N_c)$ gauge group\footnote{A study of different fermion representations can be found in, e.g., \cite{Armoni:1999xw}}.
We will mostly consider the heavy mass regime
\be
\label{heavy}
m^2\gg {g^2N_c}\;,
\ee
where one expects a straightforward perturbative treatment to apply.
The spectrum of this model was studied extensively in early 90's \cite{Dalley:1992yy,Bhanot:1993xp,Kutasov:1993gq} (see \cite{Katz:2013qua} for a recent update). A study of the worldsheet dynamics has been initiated in 
\cite{Dubovsky:2018dlk}, we will adopt the same procedure here.  

Following \cite{Coleman:1976uz,Witten:1978ka}, we put the theory on a finite interval, $\sigma\in(-L,L)$ with an infinitely heavy fundamental quark-antiquark pair $Q,\bar Q$ placed at the endpoints. Eventually,  we take the infinite volume limit $L\to\infty$. Then the worldsheet theory describes dynamics of states created by  gauge invariant single-trace operators of the form
\be
{\cal O}_w= \bar{Q}\l
P%\left\{
 e^{i\int_{\cal C}d\sigma A_\sigma}\psi(\sigma_1)\dots\psi(\sigma_N)
 %\right\}
%
 \r
 Q\;. 
\ee
Here the integration path ${\cal C}$ starts at $\sigma=-L$ and ends at $\sigma=L$, but does not have to be straight. Turning points are allowed at the locations $\sigma_i$ of the adjoint quarks insertions because the Polyakov zigzag symmetry \cite{Polyakov:1997tj} is broken in the presence of the adjoint matter. Physically, one may think that the worldsheet theory describes the interior of a heavy quark $Q\bar Q $-meson. 

After fixing the spatial gauge,
$
A_\sigma=0,
$
one can integrate out the remaining non-dynamical component $A_\tau$ of the gauge field. The resulting Hamiltonian  acts in the extended Hilbert space
\be
{\cal H}_{ex}=V\otimes{\cal H}_f\otimes \bar V\;,
\ee
where ${\cal H}_f$ is the free fermion Fock space, and $V(\bar V)$ are (anti)fundamental representations of the color group. These additional factors represent color degrees of freedom of the endpoint quarks $Q(\bar Q)$.
As a consequence of confinement, the physical  Hilbert space ${\cal H}_{ph}$ is the subspace of ${\cal H}_{ex}$ annihilated by all color charges,
\be
\label{charge}
\l {\cal T}^a+\bar {\cal T}^a+\int d\sigma \rho^a\r{\cal H}_{ph}=0\;.
\ee 
Here $\rho^a$ is the color density of adjoint fermions, and ${\cal T}^a(\bar{\cal T}^a)$ are (anti)fundamental generators representing color charges of $Q(\bar Q)$. 

To describe the physical states it is convenient to adopt operator notations following from the identification
\be
V\otimes \bar V= L(V)\;,
\ee
where $L(V)$ is the space of linear operators acting on $V$.
Then a general state in ${\cal H}_{ex}$ takes the form 
\be
|\psi\rangle=\sum_i|\psi_i\rangle_F\otimes{ M}_i\;,
\ee
with $|\psi_i\rangle_F\in {\cal H}_f$ and ${ M}_i\in L(V)$.
The physical worldsheet Hilbert space ${\cal H}_w$ is generated by color singlets  of the form
\be
\label{Hw}
{\cal H}_w=\left\{\psi^{a_1}\cdot\dots\cdot\psi^{a_N}|0\rangle_F\otimes {T}^{a_1}\cdot\dots\cdot { T}^{a_N}\right\}\;.
\ee
Here ${ T}^a$'s are fundamental  $SU(N_c)$ generators considered as elements of $L(V)$. These are not to be confused with the quantum operators ${\cal T}^a$'s in (\ref{charge}). Note that ${\cal H}_{ph}$ contains additional 
multitrace color singlet states such as the mesonic state $\psi^a\psi^a|0\rangle_F\otimes {\mathbb 1}$. Multitrace states decouple from the worldsheet in the planar limit.

In the momentum representation the physical worldsheet states can be written as linear combinations of
\be
\label{ks}
|k_1,\dots,k_N\rangle= {1\over \sqrt{2}}\l{2\over N_c}\r^{N+1\over 2}
%{2^{n\over2}\over N_c^{n+1\over2}}
%b^{a_1\dagger}_{k_1}\dots b^{a_n\dagger}_{k_n}
\prod_{i=1}^Nb^{a_i\dagger}_{k_i}
|0\rangle_F\otimes 
\prod_{i=1}^N T^{a_i},
%T^{a_1}\dots T^{a_n}\;.
\ee
where $b_k^{a\dagger}$ are fermionic creation operators.
For many purposes it is convenient to also use  the coordinate representation. The corresponding basis is
\be
\label{sigmas}
|\sigma_1,\dots,\sigma_N\rangle=\int\prod_{i=1}^N{dk_i\over\sqrt{2\pi}}e^{-ik_i\sigma_i}|k_1,\dots,k_N\rangle\;.
\ee
In the heavy mass regime (\ref{heavy}) eigenstates of the worldsheet Hamiltonian are well characterized by their parton number. In particular, one finds the worldsheet vacuum  
$
|0\rangle=|0\rangle_F\otimes {\mathbb 1}
$
and one-particle excitations 
$
|k\rangle={\sqrt{2}\over N_c}b^{a\dagger}_{k}\otimes T^a,
$
 dubbed ``free quarks" in \cite{Witten:1978ka}.

As emphasized in \cite{Dubovsky:2018dlk}, the worldsheet multiparticle  states (\ref{ks}), (\ref{sigmas}) do not describe conventional identical particles. For instance, the configuration space of two-particle states (\ref{sigmas}) is the whole plane instead of the half-plane, as $|\sigma_1,\sigma_2\rangle\neq|\sigma_2,\sigma_1\rangle$.
% as one might had expected given that the worldsheet dynamics is described by a Lorentz invariant quantum theory. 
 Equivalently, in the planar limit the exchange term is missing in multiparticle inner products,
\be
\label{inner}
\langle\sigma'_N,\dots,\sigma'_1|\sigma_1,\dots,\sigma_N\rangle=\prod_{i=1}^N\delta(\sigma_i-\sigma'_i)\;.
\ee
This is the inner product for a system of $N$ distinguishable particles, indicating that the worldsheet theory provides a non-trivial dynamical realization of infinite quon statistics  (see, e.g., \cite{Greenberg:1989ty}).
It also serves as an interesting counterexample to the common lore \cite{Weinberg:1995mt} that the conventional Fock space is the only possible arena for Lorentz invariant quantum dynamics.

Up to now our treatment of the worldsheet theory was mostly kinematical. To study dynamics and to implement the program outlined in the introduction we need to evaluate how the worldsheet Hamiltonian acts on the physical space 
${\cal H}_w$ and to learn how to develop perturbation theory in this unconventional Hilbert space.  We leave this task for a separate publication \cite{future} and will discuss here only the very first step---identification of an unperturbed Hamiltonian $ H$. %This will already yield quite interesting and surprising results.

%In principle, it is straightforward to evaluate matrix elements of the full worldsheet Hamiltonian between the physical worldsheet states (\ref{ks}) or (\ref{sigmas}) generalizing the calculation presented in  \cite{Dubovsky:2018dlk}
%for two-particle matrix elements. The result is somewhat cumbersome and we don't actually need it here. Instead, l
Let us first inspect two-particle matrix elements of the full worldsheet Hamiltonian $H_w$. In the c.o.m. frame one finds \cite{Dubovsky:2018dlk}
\[
\langle k_2,k_1|H_w|k,-k\rangle=\delta(k_1+k_2)
\l \delta(k_1-k)2\omega_k-{g^2N_c\over 4\pi} {\cal V}
\r.
\]
Here $\omega_k=\sqrt{k^2+m^2}$ and
\be
\label{calV}
{\cal V}={{\cal U}(k,k_1)}{{\cal P}\over (k-k_1)^2}+ i\pi\delta'(k-k_1)-{m^2\over 4\omega_{k}^2\omega_{k_1}^2}\;,
\ee
where ${\cal U}(k,k_1)$ is a smooth function of momenta. It is equal to unity for $k,k_1\ll m$, or $k,k_1\gg m$ and also in the forward limit $k=k_1$.
${\cal P}$ stands for the principal value.

Even though ${\cal V}$ is multiplied by the 't Hooft coupling, it cannot be entirely treated as a perturbation due to the presence of forward singularities in (\ref{calV}). The physical meaning of these singularities is transparent in  position space. Setting ${\cal U}(k,k_1)=1$, dropping the last non-singular term in  (\ref{calV}) and switching to position space we arrive at
\be
\label{V0}
{\cal V}_0(\sigma)=\sigma+|\sigma|\;.
\ee
We see that the conventional statistics gets restored on-shell---the growth of the potential (\ref{V0}) at $\sigma=+\infty$ reduces the number of scattering states by a factor of two, restoring the agreement with  state counting for conventional identical particles. 

It is straightforward to check that the story repeats for multiparticle states. Namely,  all terms in the full Hamiltonian which grow at spatial infinity are diagonal 
in the particle number and  combine into  the following potential in the $N$-particle sector
\be
\label{Vn}
V_N={g^2N_c\over 4\pi}\sum_{i=1}^{N-1}{\cal V}_0(\sigma_{i,i+1})\;,
\ee 
where 
$
\sigma_{i,i+1}=\sigma_i-\sigma_{i+1}.
$
As a result, in the asymptotic regions $\tau\to\pm\infty$ one only finds  configurations with $\sigma_1\leq\sigma_2\dots\leq\sigma_N$ and
the conventional statistics gets restored on-shell for any number of colliding particles. Zigzags responsible for the emergence of the off-shell infinite statistics cost energy and do not survive on-shell.

%\section{Integrable Worldsheet Mechanics}
{\it Integrable Worldsheet Mechanics.}
These results  suggest the following choice of the unperturbed Hamiltonian in the worldsheet perturbation theory,
\be
\label{Hn}
 H_{N,m}=\sum_{i=1}^N\sqrt{p_i^2+m^2}+V_N\;,
\ee
where the subscript $N$ indicates that (\ref{Hn}) is a restriction of the unperturbed Hamiltonian $ H$ to the $N$-particle sector. In addition to the conventional free piece, $H_{N,m}$  also incorporates the leading forward singularities, which determine the asymptotic growth of the potential. A similar leading order Hamiltonian was also obtained  in the Abelian case (at $N=2$) based on $\hbar$ counting \cite{Coleman:1976uz}.

To see whether (\ref{Hn}) is a good choice it is natural to check at least the following two conditions:
\\
{\it (i)} Is   (\ref{Hn})  Poincar\'e invariant?
 \\
 {\it  (ii)} Is   (\ref{Hn})  integrable?
 
We restrict to the classical analysis of (\ref{Hn}). Even though the procedure which lead us to (\ref{Hn}) was not manifestly Lorentz covariant, the final result is. Indeed, piecewise the potential (\ref{Vn}) is  either free or describes a constant electric field acting on some of the particles. Both options correspond to Lorentz invariant dynamics in two dimensions. More precisely, a single particle moving in a constant electric field is invariant under the centrally extended Poincar\'e group \cite{Karat:1998qg}. However, the central charge vanishes for (\ref{Hn}) after contributions from all particles are added up. The corresponding boost generator is
\be
J=\sum_{i=1}^N \sigma_i\sqrt{p_i^2+m^2}+{1\over 2}\sum_{i=1}^{N-1}(\sigma_i+\sigma_{i+1}){\cal V}_0(\sigma_{i,i+1})\;.
\ee
The Poisson brackets between $ H$, $J$ and the total momentum $P$ satisfy the $ISO(1,1)$ Poincar\'e algebra
\be
\label{HPJ}
\{ H,P\}=0\;,\;\;\{J,P\}= H\;,\;\;\{J, H\}=P\;.
\ee
One might expect  then that the system (\ref{Hn}) is also integrable, {\it i.e.} no momentum transfer occurs in multiparticle collisions. Indeed, as a consequence of crossing symmetry, momentum transfer 
implies particle production. The latter is absent for $ H$. In accord with this argument, in the relativistic two-dimensional\footnote{Recall that at $D>2$ the ``no interaction theorem" \cite{Currie:1963rw,Currie,Leutwyler1965} excludes 
 interacting finite-dimensional relativistic Hamiltonian systems.
} $N$-body systems of \cite{Ruijsenaars:1986vq,ruijsenaars1990} integrability automatically follows from  Poincar\'e invariance.

However, a straightforward {\it Mathematica} simulation of (\ref{Hn}) shows that the system is {\it not} integrable. Most likely this is related to difficulties with preserving the Poincar\'e algebra (\ref{HPJ}) at the quantum level in very similar models \cite{Lenz:1995tj,Kalashnikova:1997cs}. We leave it to \cite{future} to study whether this may be resolved by including  subleading forward singularities  in $ H$. Instead, here we observe that the high energy limit of (\ref{Hn}) given by 
\be
\label{Hn0}
H_N\equiv H_{N,0}=\sum_{i=1}^N|p_i|+\sum_{i=1}^{N-1}\l \sigma_{i,i+1}+|\sigma_{i,i+1}|\r\;,
\ee
does give rise to a Poincar\'e invariant integrable $N$-body model. From now on we set the 't Hooft coupling to unity, 
$
{g^2N_c=4\pi}$. Note that this limit does not contradict (\ref{heavy}), as $p^2\gg m^2\gg g^2N_c$.
%We used (\ref{heavy}) to arrive at (\ref{Hn}) and then neglected $m^2$ compared to $p_i^2$. 
On the other hand, (\ref{Hn0}) does not actually rely on  (\ref{heavy})---this Hamiltonian describes the high energy worldsheet dynamics independently of whether (\ref{heavy}) holds or not.

\begin{figure*}[th!]
  \begin{center}
        \includegraphics[width=6cm]{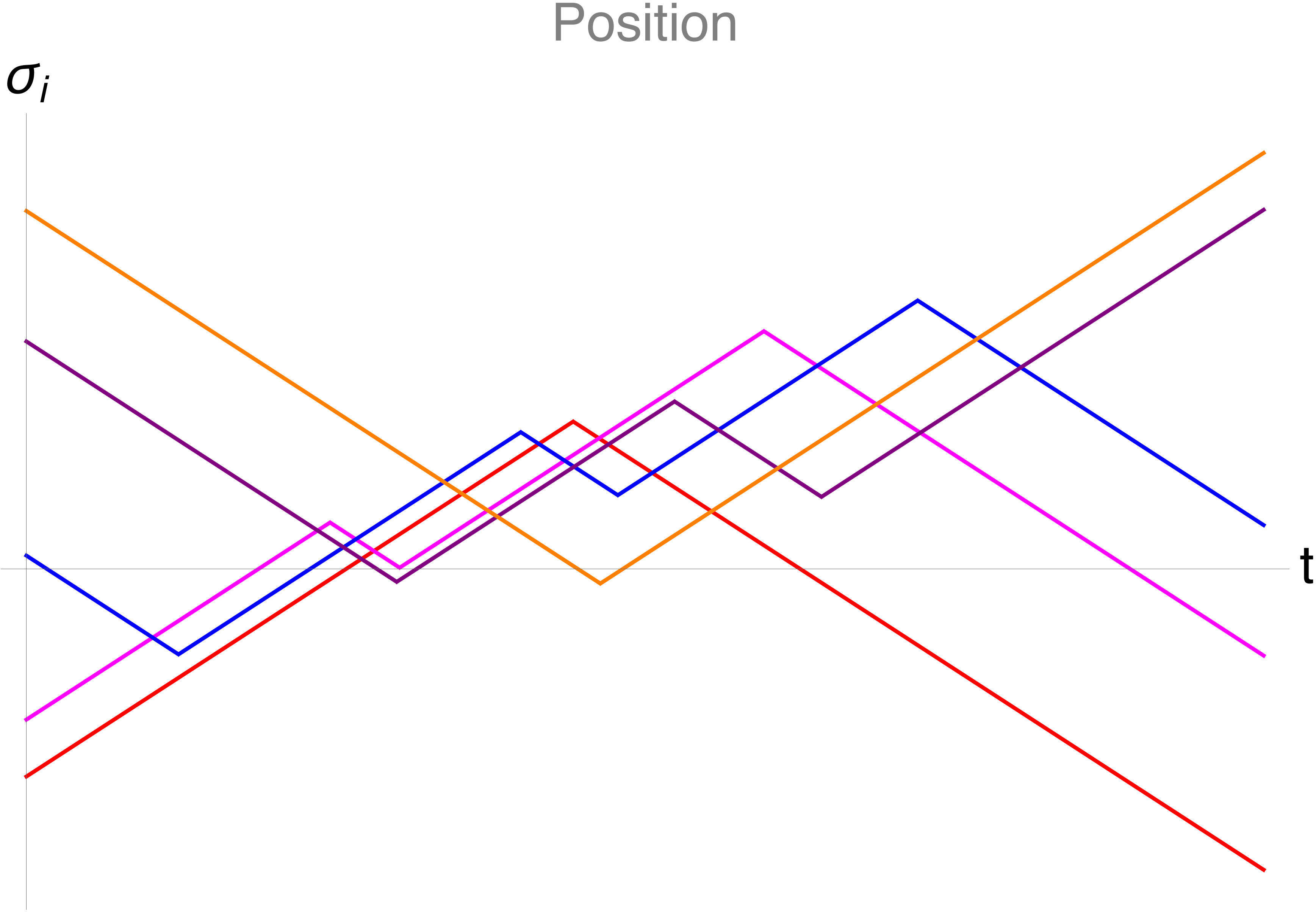} 
        \hskip2cm
         \includegraphics[width=6cm]{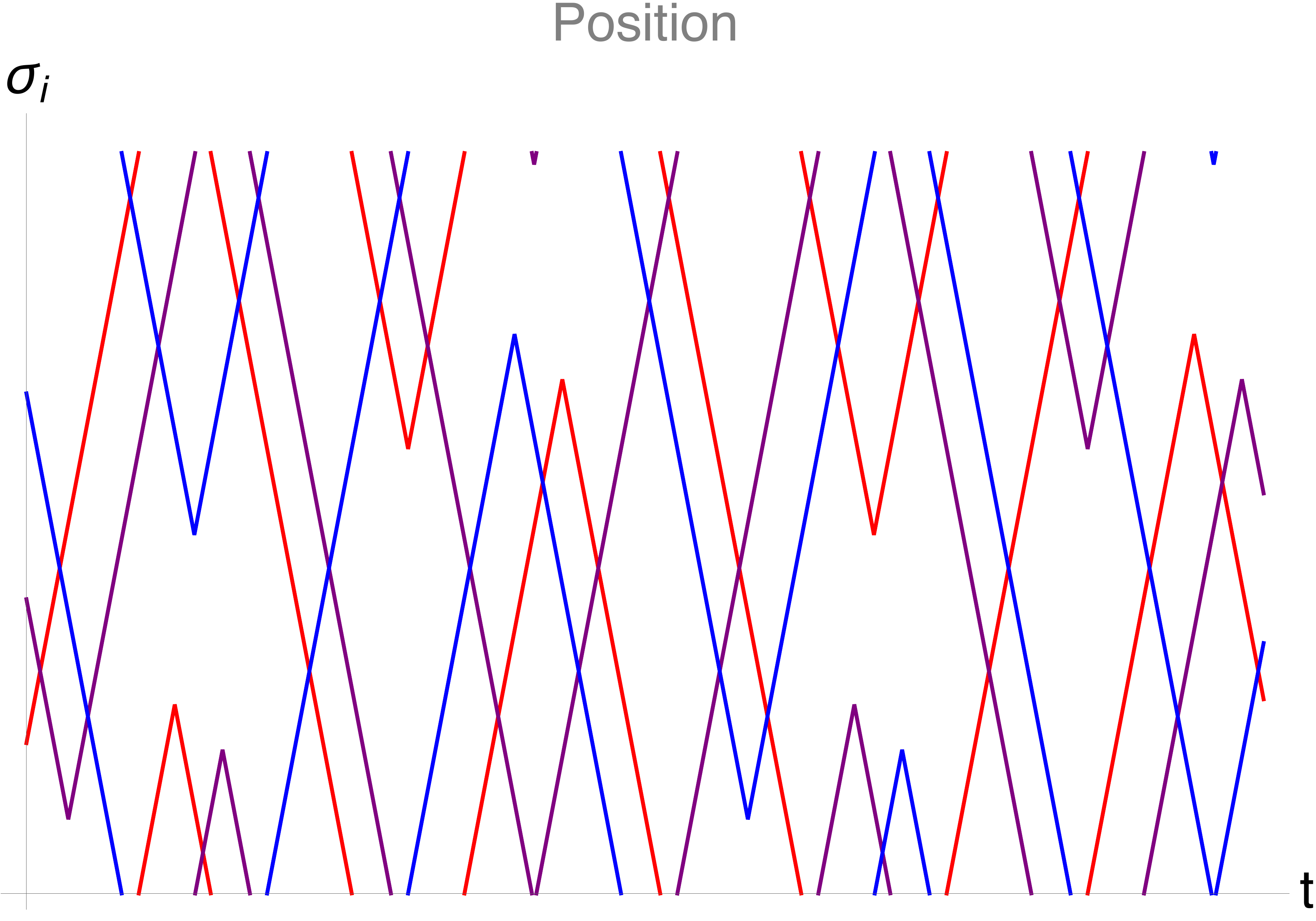} 
         \includegraphics[width=6cm]{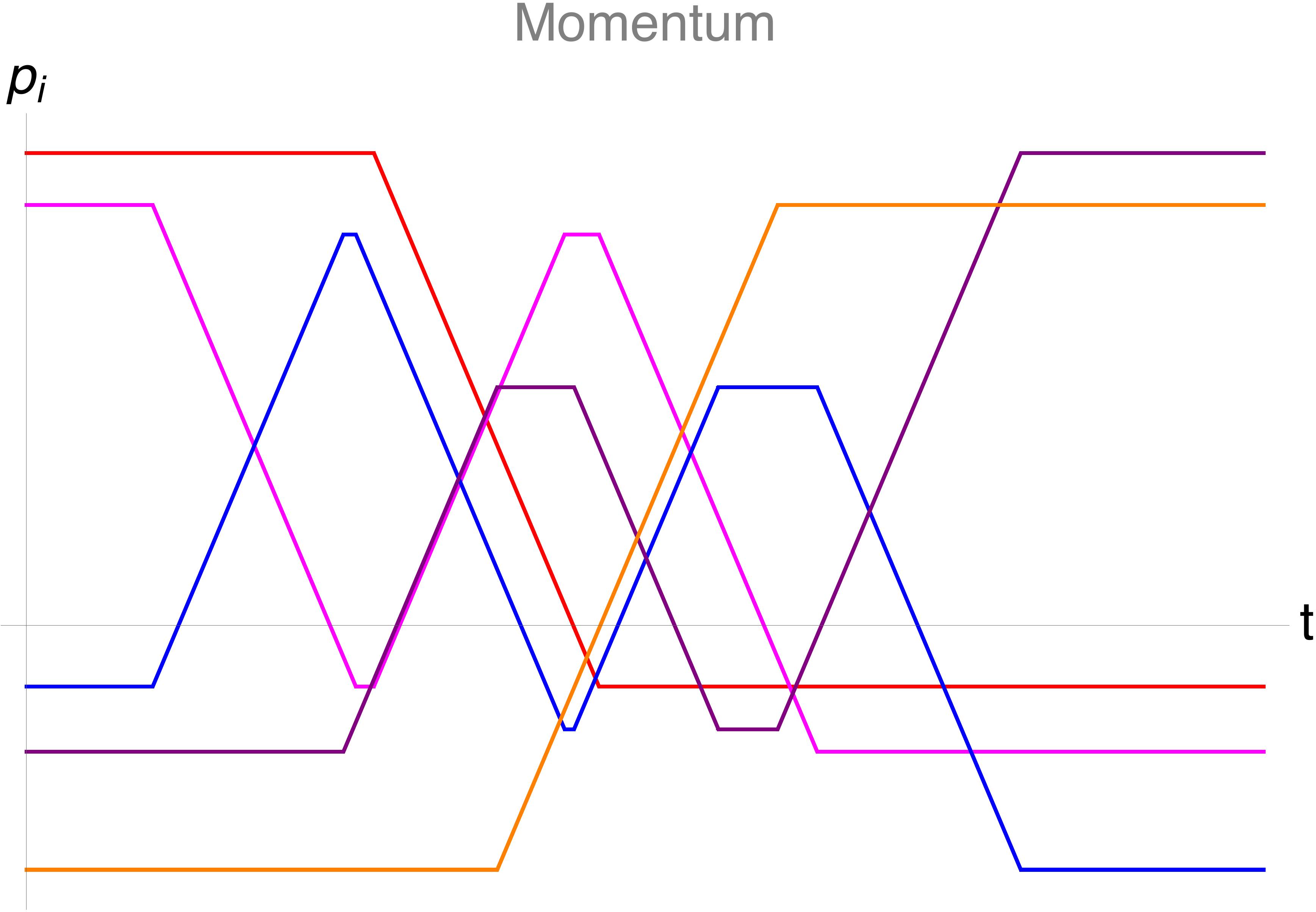}  
                 \hskip2cm
        \includegraphics[width=6cm]{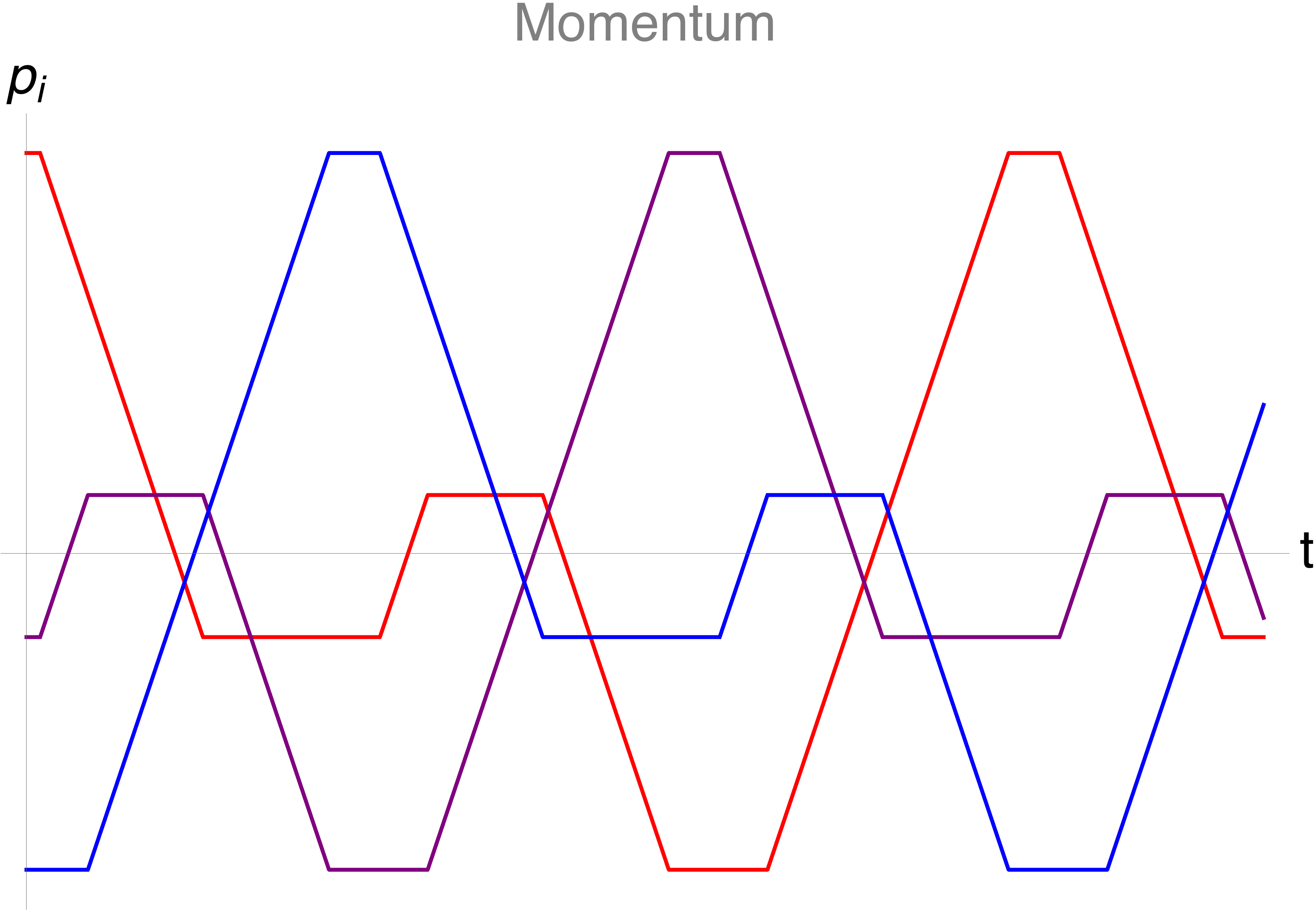} 
           \caption{ Time evolution for a sample 5-particle scattering (left) and of a  3-particle configuration in the periodic case (right).}
        \label{fig:sols}
    \end{center}
\end{figure*}

Integrability of (\ref{Hn0}) is the main observation of this paper. It confirms that (\ref{Hn0}) does provide a good starting point for the high energy expansion on the worldsheet.
The fastest way to check integrability is to solve the equations of motion,
\be
\label{eqs}
\dot\sigma_i=s_i\;,\;\;\dot p_{i}=s_{i-1,i}-s_{i,i+1}\;.
\ee
Here $s_i=\mbox{sign\,}p_i$, $s_{0,1}=s_{N,N+1}=-1$ and %=1,\dots,N-1$ 
$s_{j,j+1}=\sign\sigma_{j,j+1}$ for $1\leq j<N$. A general solution of (\ref{eqs})  is a piecewise linear function of time.
% $\tau$,
%\be
%\label{sol}
%\sigma_i=a_i(S)+s_i\tau\;,\;\;p_i=b_i(S)+(s_{i-1,i}-s_{i,i+1})\tau\;,
%\ee
%where  $S$ contains all $s_i$'s and $s_{i,i+1}$ with $1\leq i<N$.%=1,\dots,N-1$.
Using  {\it Mathematica}\footnote{An exact {\it Mathematica} solver and the resulting movies can be downloaded at \url{https://jcdonahue.net/research}.} it does not take long to convince oneself
 that the  initial and final asymptotic sets of momenta are always the same. As an illustration, in Fig.~\ref{fig:sols} we plot a non-trivial five-particle collision and  a self-repeating three-particle solution of the integrable periodic version of the model obtained by setting $s_{0,1}=s_{N,N+1}=\sign(\sigma_N-\sigma_1-2\pi)$.

We will present a detailed study of the resulting integrable structure in \cite{future} and restrict  here to just a few  remarks.
Given the piecewise linear time dependence of the solutions, it is natural to look for conserved topological invariants $T(S)$ and also for stepwise linear conserved charges of the form
\be
\label{genI}
I(\sigma,p)=A^i(S)\sigma_i+B^i(S)p_i\;,
\ee
where   $S=(s_{0,1},s_1,\dots,s_n,s_{n,n+1})$ is the set of all signs.
%stand for all phase space coordinates. 
 % Instead, these are higher order polynomials in $S$.
For instance, using (\ref{eqs}) it is straightforward to check that  
\be
\label{I2}
T_{2}={1\over 2}\sum_{i=1}^{N} s_i(s_{i-1,i}+s_{i,i+1})
\ee
 stays constant. To see the physical meaning of $T_2$ let us evaluate it in the asymptotic regions, where all $s_{i-1,i}=-1$.
One finds,
\[
T_{2}=N_L-N_R\;,
\]
where $N_{L(R)}$ are the numbers of left(right)-movers in the asymptotic regions. It is natural to think of  $S$ as a sequence of classical spin variables. Interestingly, $T_2$ turns then into the Ising Hamiltonian.

$T_2$ also provides a convenient starting point to construct  a non-topological charge  by looking for $F_{T_2}$ satisfying
$
%\label{FT}
\{F_{T_2},H\}=T_2.
$
Using (\ref{eqs}) it is straightforward to construct $F_{T_2}$ in the non-periodic case,
\[
F_{T_2}=\sum_{k=1}^N\l k-{1\over 2}\r |p_k|+\sum_{k=1}^{N-1}k|\sigma_{k,k+1}|-N\l\sigma_N+{H_N\over2}\r.
\]
$F_{T_2}$ itself is only conserved in the sector with $N_L=N_R$. However, it is possible to build a new conserved quantity in all sectors by acting on $F_{T_2}$ with the boost generator.
Namely, let us define
\[
\tilde{P}=T_2\{J,F_{T_2}\}-N F_{T_2}%\l F_{T_2}-{N\over 2}H\r-T_2 \{J,F_{T_2}-{N\over 2}H\}
\;,\;\;
\tilde{H}=T_2F_{T_2}-N \{J,F_{T_2}\}% {N\over 2}H\r-N \{J,F_{T_2}-{N\over 2}H\}
\;.
\]
Then the Poincar\'e algebra (\ref{HPJ}) gets enlarged to 
\begin{gather}
\label{HPJt}
\{\tilde H,\tilde P\}=0\;,\;\;\{J,\tilde P\}=\tilde H\;,\;\;\{J,\tilde H\}=\tilde P\\
\{H,\tilde H\}=\{\tilde{P}, P\}=N^2-T_2^2\;,\;\;\{P,\tilde H\}=\{H,\tilde P\}=0\nonumber\;.
\end{gather}
 It was suggested in \cite{Dubovsky:2018dlk} that in a putative gravitational description of the worldsheet the zigzags should correspond to black holes and the exotic off-shell statistics  to black hole complementarity. In this context it is encouraging to find the ``shadow" Poincar\'e subalgebra $(\tilde H,\tilde P, J)$. It will be interesting to check whether the shadow charges $(\tilde H,\tilde P)$ may be identified  with the Hamiltonian and momentum seen by infalling observers.

Similar techniques also allow one to obtain additional integrals in involution at $N>2$. For instance, for $N=3$  the following translationally invariant  combination is conserved in the $N_R=2$, $N_L=1$ sector
\[
%I_{12}=p_1+|p_1|+\sigma_{1,2}+|\sigma_{1,2}|+{1\over 4}(p_2+|p_2|)(3+s_1(s_{1,2}-1)+s_{1,2})\;.
I_{12}={\cal V}_0(p_1)+{\cal V}_0(\sigma_{1,2})+{{\cal V}_0(p_2)\over 4}(3+s_1(s_{1,2}-1)+s_{1,2}).
%p_1(1+s_1)+\sigma_{1,2}(1+s_{1,2})+{p_2\over 4}(1+s_2)(3+s_1(s_{1,2}-1)+s_{1,2}).
\]
This establishes  Liouville integrability at $N=3$. Interestingly, $I_{12}$ is {\it not}
a higher order polynomial in  momenta, as typically found using the Lax pair technique in similar integrable models (such as the Toda chain~\cite{Bazhanov_2017}). We expect that also at $N>3$ the ansatz (\ref{genI}) leads 
 to $N-2$   additional independent charges in involution.

%\section{Future Directions}
{\it Future Directions and Relation to $T\bar T$.}
The presented results open numerous avenues for  future research. In particular, what about the quantum integrability of (\ref{Hn0})? The following argument indicates that it should  hold. The classical two-particle phase shift
following from (\ref{Hn0}) is $\delta=s$, which coincides with the exact phase shift of a known quantum integrable model---the $T\bar T$ deformation of a free massless fermion \cite{Zamolodchikov:2004ce,Dubovsky:2012wk,Dubovsky:2013ira,Smirnov:2016lqw,Cavaglia:2016oda} (see, e.g., \cite{Bonelli:2018kik,Coleman:2019dvf} for explicit derivations of the corresponding classical action). This indicates that there exists a quantization of (\ref{Hn0}) resulting in the same phase shift, and that  (\ref{Hn0}) describes the $N$-particle subsector of the $T\bar T$ deformed fermion, similarly to how the Ruijsenaars--Schneider model describes the $N$-particle subsector of the sine-Gordon theory \cite{Ruijsenaars:1986vq}. The appearance of the shadow Poincar\'e subalgebra in (\ref{HPJt}) mirrors the presence of dynamical and worldsheet ``clocks and rods" in the gravitational formulation of the $T\bar T$ deformation \cite{Dubovsky:2017cnj,Cardy:2018sdv,Dubovsky:2018bmo,Conti:2018tca}.
 This relation also suggests that the infinite quon Hilbert space ${\cal H}_w$ may provide a natural arena for defining off-shell observables in $T\bar T$ deformed theories. It will be interesting to  connect this to the recent construction of \cite{Cardy:2019qao}. Finally, this relation indicates that an integrable relativistic $N$-body system described here is actually a model of dynamical geometry (gravity), see also \cite{KITP}.
 
 As far as the $D=2$ QCD physics goes, it was noticed in numerical studies of the spectrum that meson mass eigenstates have definite parton numbers with a very high accuracy even for small quark masses \cite{Dalley:1992yy,Bhanot:1993xp}.   It was also  observed that the equation for the spectrum becomes exactly solvable in the high energy limit \cite{Kutasov:1993gq}. Both these observations should be related to the integrability found here and it will be useful to make the connection precise.
 
 In addition, it will be very interesting to see which of  the presented results can be generalized to $D=3,4$ and to connect this approach to other signs of approximate integrability in QCD, such as \cite{Ferretti:2004ba,Beisert:2004fv}. It is encouraging to see that an integrable $T\bar{T}$-deformed model emerged at high energies in the $D=2$ case, given that according to the ASA the integrable models appearing at $D=3,4$ are also $T\bar{T}$ deformations.
%\section*{Acknowledgements}

{\it Acknowledgements.}
We thank  Misha Feigin, Vitya Gorbenko, Sam Grushevsky, Misha Ivanov, Igor Klebanov, Zohar Komargodski, Grisha Korchemsky, Conghuan Luo and Riccardo Rattazzi for useful discussions. 
This work is supported in part by the NSF CAREER award PHY-1352119.
\bibliographystyle{utphys}
\bibliography{dlrrefs}
\end{document}